\documentclass{aa}  

\usepackage{graphicx}
\usepackage{float}
\usepackage{txfonts}
 \usepackage[normalem]{ulem}
 \useunder{\uline}{\ul}{}
\usepackage{multirow}

\usepackage[colorlinks=true, allcolors=blue]{hyperref}
\begin{document} 

   \title{Assessing the processes behind planet engulfment and its imprints}


   \author{B. M. T. B. Soares
          \inst{1,2}
          \and
          V. Adibekyan\inst{2}
          \and
          C. Mordasini\inst{3}
          \and
          M. Deal \inst{4}
          \and
          S. G. Sousa \inst{2}
          \and
          E. Delgado-Mena\inst{2}
          \and
          N. C. Santos\inst{1,2}
          \and
          C. Dorn \inst{5}
          }

   \institute{Departamento De Física e Astronomia, Faculdade de Ciências, Universidade do Porto, rua do Campo Alegre s/n, 4169– 007 Porto, Portugal,
              \\
              \email{Barbara.Soares@astro.up.pt}
         \and
             Instituto de Astrofísica e Ciências do Espaço, Centro de Astrofísica da Universidade do Porto, Rua das Estrelas 4150-762 Porto, Portugal
         \and
             Institute of Physics, University of Bern, Gesellschaftsstrasse 6, 3012 Bern, Switzerland
         \and
             LUPM, Universit\'{e} de Montpellier, CNRS, Place Eug\`{e}ne Bataillon, 34095 Montpellier, France            
        \and
             Institute for Particle Physics and Astrophysics, ETH Zürich, Otto-Stern-Weg 5, 8093 Zürich, Switzerland
             }

   \date{}

 
  \abstract
   {Newly formed stars are surrounded by a protoplanetary disk composed of gas and dust, part of which ends up forming planets. During the system's evolution, some of the planetary material may end up falling into the host star and be engulfed by it, leading to a potential variation of the stellar composition.} 
   {The present study explores how planet engulfment may impact the chemical composition of the stellar surface and discusses what would be the rate of events with an observable imprint, for Sun-like stars.}
   {We use data on the formation and evolution of 1000 planetary systems from the New Generation Planetary Population Synthesis (NGPPS) calculations by the Generation III Bern model to analyse the conditions under which planet engulfment may occur. Additionally, we use stellar models computed with Cesam2k20 (Code d’Evolution Stellaire Adaptatif et Modulaire) to account for how the stellar internal structure and its processes may affect the dilution of the signal caused by planet engulfment.}
   {Our results show that there are three different phases associated to different mechanisms under which engulfment events may happen. Moreover, systems that undergo planet engulfment are more likely to come from protoplanetary disks that are more massive and more metal-rich than non-engulfing systems. Engulfment events leading to an observable signal happen after the dissipation of the protoplanetary disk when the convective envelope of the stars becomes thinner.
   With the stellar convective layer shrinking as the star evolves in the main sequence, they display a higher variation of chemical composition. This variation also correlates with the amount of engulfed material. By accounting for the physical processes happening in the stellar interior and in the optimistic case of being able to detect variations above 0.02 dex in the stellar composition, we find an engulfment rate no higher than $20\%$ for Sun-like stars that may reveal detectable traces of planet engulfment.}
   {
   Engulfment events that lead to an observable variation of the stellar composition are rare due to the specific conditions required to result in such signatures. 
}


   \maketitle
%

\section{Introduction} \label{intro}

There is a rich variety among the more than 5700 exoplanets\footnote{As per \href{exoplanet.eu}{The Extrasolar Planets Encyclopaedia}, consulted on 04/06/2024.} detected, and they can range from super-Mercuries (e.g. \citealt{2022A&A...657A..68A}) to hot Jupiters (e.g. \citealt{2022AJ....163..120G})
and even deformed planets \citep{2022A&A...657A..52B}. 
The diversity of the exoplanet population challenges our understanding of the formation and evolution of planetary systems.

Given that stars and planets form from the same initial gas and dust cloud, it is common to assume that the composition of planet building blocks should be related with the composition of the stellar atmosphere (e.g. \citealt{Asplund2009}). The wide range of stellar compositions (e.g. \citealt{2012A&A...545A..32A}) is believed to be reflected in the composition of rocky planets (e.g. \citealt{2017A&A...608A..94S}), directly affecting the bulk densities of planets.
In particular, the correlation between the composition of planet building blocks and the stellar atmosphere is valid for refractory elements \citep{bond2010,thi2014,bonsor2021}.
As a matter of fact, it has been shown that the knowledge of the relative abundances of refractory major rock-forming elements such as Fe, Mg and Si contributes to the improvement of interior estimates for telluric planets \citep{dorn2015,Unterborn_2016}.
Although the composition of terrestrial planets and their host stars was always assumed to correlate through a 1-to-1 linear relation, \cite{2021Sci...374..330A} recently demonstrated that this assumption is not valid.

Through the in-falling of planets and their subsequent absorption into their host star, the evolution of planets can have a measurable impact in the stellar surface chemical composition. The pioneering work of \cite{1997MNRAS.285..403G} initiated the search for signatures of planet formation and engulfment in the composition of solar-like stars resulting, however, in contradictory results (\citealt{2001ApJ...556L..59P,2009A&A...508L..17R,2010A&A...521A..33R,Hernandez_2010}, \citeyear{2011IAUS..276..422G}, \citeyear{2013A&A...552A...6G}, \citealt{2017AN....338..442A}, \citealt{2021MNRAS.508.1227L}).

The work by \cite{2009ApJ...704L..66M} has shown an anomaly in the solar chemical composition when compared to other solar-twins that they interpreted as due to a deficit in refractories relative to volatiles. Alternatively, \cite{2022MNRAS.512.3684C} explore the option of the Sun being instead enhanced in volatiles, but they find that the best quality data set does not support this hypothesis.
To explain this anomaly, \cite{2009ApJ...704L..66M} suggested that the observed depletion in refractory elements is a marker of planet formation, since the formation and retention of solid planetary material leads to the star accreting disc material depleted in certain heavy elements.
However, \cite{2020MNRAS.493.5079B} argue that this deficit does not come from retention of refractory material inside planets. Instead, it is likely to appear from gaps in protoplanetary disk due to the formation of Jupiter analogues and consequent trapping of dust outside of their orbits, which prevents it from being accreted into the host star. They find that the forming giant planets can create refractory depletion of $\sim$ 5 - 15$\%$, with initial conditions favouring giant planet formation leading to the larger values of refractory depletion.
In their review paper, \cite{2017AN....338..442A} present two alternative scenarios that could explain this irregularity in the solar chemical composition. The chemical anomaly could either just reflect the galactic chemical evolution and be related with the stellar age \citep{2014A&A...564L..15A}, or it could be due to the fact that the Sun formed in dense stellar environment, in a very massive open cluster \citep{2014A&A...562A.102O}.

The difference in chemical composition could, however, be seen in another perspective. Instead of the Sun being depleted in refractories relative to volatiles, it could be the solar-twins being enriched in refractory elements  \cite[e.g.][]{2014A&A...567A..55S,2015A&A...582L...6S}. \cite{smith2001} and \cite{2011ApJ...732...55S} proposed that the accretion of metal-rich planetary material onto the star can result in an enhancement in heavy elements of the stellar surface composition through the infall of solid planets or minor bodies, i.e. planet engulfment \citep{2011ApJ...740...76R,2015A&A...582L...6S,2020MNRAS.491.2391C}. Therefore, it is not easy to ascertain the physical reason behind the differences in the chemical composition of solar-like stars and the Sun.

Binary stars are the best laboratories to test the impact of planet engulfment on the host star composition, as stars from binary systems are formed from the same molecular cloud, and thus should possess similar chemical compositions \citep{2018ASSP...49..225A}. This has been done on individual system basis, for binaries where one of the pairs has a planetary companion. There are many contradictory results and while some reveal the planet hosting binary component showing a significant enhancement in elemental refractory abundances \cite[e.g.][]{2017A&A...604L...4S,2018ApJ...854..138O,2019MNRAS.490.2448R,2020ApJ...888L...9N},
others find no such difference \cite[see e.g.][]{2014MNRAS.442L..51L,2015A&A...582A..17S,2016ApJ...818...54M}. For the cases where there is a refractory enhancement in the chemical composition, the authors attribute these differences in abundances of 0.1 - 0.2 dex to consequences of planet engulfment.
Nonetheless \cite{2015A&A...584A.105D} , \cite{2019MNRAS.490.2448R} and \cite{2021MNRAS.508.1227L} argue that offsets observed in binary systems could be due to atomic diffusion and its consequences. Furthermore, both find a weak correlation between binary separation and chemical abundance differences. This result was recently explored by \cite{2023MNRAS.521.2969B,2023MNRAS.518.5465B}, who find that differences in abundances for stars within a wide binary may be primordial, since large projected separations can exceed the typical turbulence scales in molecular clouds, and thus have no connection to the presence of planets.

Variation of stellar composition due to planet engulfment was once more brought up in a recent work by \citealt{2021NatAs...5.1163S} (hereinafter referred to as SP21), who analysed a sample of 107 binaries composed by pairs of Sun-like stars, and reported how the binaries components had different compositions.
The authors attributed this difference to the engulfment of terrestrial material into one of the binary components, and claimed an engulfment rate of $\sim 20 - 35\%$ for their sample. Moreover, SP21 suggested the engulfment should have occurred after the disk dissipation, as otherwise the impact of the engulfment event would mostly disappear due to the large size of the convective zone of the star.

Nevertheless, \cite{2023MNRAS.521.2969B,2023MNRAS.518.5465B} suggest that detection of planet engulfment signatures are difficult to detect in solar-like stars. Thermohaline mixing and gravitational settling \citep{2009ApJ...704.1262T,2012ApJ...753...49V,2015A&A...584A.105D} weaken engulfment signatures, leading to a signal observable only up to $\sim 2$ Gyr after the engulfment event, depending on the mass, temperature and metallicity of the star. \citeauthor{2023MNRAS.521.2969B} conclude that engulfment is hard to detect in systems that are several Gyr old, and argue that the true engulfment rate announced by SP21 is closer to $\sim 2.9\%$ instead, a value comparable to results from models.

The orbital architecture of a system also plays a role in the occurrence of engulfment events. In the absence of a disk, dynamical instabilities in the system can also lead to these events. Phenomena such as fly-by encounters \citep{2014prpl.conf..787D} can disturb the system, leading to instabilities within it on long timescales.
However, it is not possible to observe the planet engulfment phase, as we can only see the current configuration of a system \citep{2023MNRAS.518.5465B,2023MNRAS.521.2969B}. As such, it is necessary to resort to models, seeing that simulations on the formation and evolution of planetary systems provide the necessary information on the state of a system across time.

In this work, we aim at understanding what is the planet engulfment rate, under which conditions do these events happen, how it correlates with the properties of the planetary systems and how it impacts the composition of the stellar surface. This is done using simulated planetary systems from formation models.
In Section \ref{data} we describe the planetary population synthetic data set we used in this work and its importance for the topic at hand. 
We start by presenting a more statistical approach of the results in Section \ref{results}, with a deeper analysis of parameters impacting planet engulfment being addressed in Section \ref{parameters}. This is followed by an analysis of the chemical composition of the planetary material engulfed by the host star and that of the remaining planets in the systems in Section \ref{variation composition}. We then examine in Section \ref{morgan} how planet engulfment events impact the stellar surface chemical composition. Section \ref{discussion} features a discussion of the results, with a final summary being provided in Section \ref{conclusions}.

\section{Data} \label{data}

To study planetary engulfment and better understand what may lead to it, we analyse the latest data from the New Generation Planetary Population Synthesis (NGPPS) calculations by the Generation III Bern model \citep{ngpps1}. The Bern model is a self-consistent global model following the core accretion paradigm \citep{1980PThPh..64..544M} and based on previously existing models built to simulate a giant planet formation in the Solar system \citep{alibert2004,alibert2005}.
It was then improved to include quantitative statistical comparisons with observations \citep{mordasini2009}, to follow the long-term evolution of the formed planet \citep{mordasini12a,mordasini12b}, and to address the structure of the protoplanetary disk \citep{fouchet2012} and the solids accretion rate \citep{fortier2013}. Additionally the model incorporates a planet migration scheme with Type I \citep{paar2011} and Type II \citep{ditt2014} migration. 
The model was then extended to allow the formation of multi-planetary systems via N-body integration \citep{alibert2013}, to include the cooling, contraction, envelope evaporation and general and thermodynamic evolution after the planetary formation over Gyr timescales \citep{mordasini12b}, and to account for the atmospheric escape \citep{jin2014}.

The data set used includes the simulation of the formation and evolution of 1000 planetary systems across 10 Gyr. Each synthetic system initially contains 100 embryos of $0.01\ M_\oplus$ orbiting a $1\ M_\odot$ star. The stellar metallicity comes from a normal distribution $\mathcal{N}(-0.02, 0.22)$ dex truncated to $-0.6 <$ [Fe/H] $< 0.5$ dex to avoid metallicities not occurring in the solar neighbourhood.
All systems initially possess a protoplanetary disk, whose gas mass comes from a log-normal distribution $\mathrm{Lognormal}(-1.49, 0.35)\ M_\odot$ constrained to $4\times 10^{-3} < M_{gas\ disk} < 0.16\ M_\odot$  to ensure the disks are always self-gravitationally stable (see \citealt{ngpps2} for a comprehensive explanation on the initial conditions of the simulations and respective distribution).

The embryos present in the systems grow (via the accretion of planetesimals and gas), collide with each other in giant impacts, migrate, and interact via N-body in an evolving protoplanetary disk. The model is divided in two phases and takes place across 10 Gyr: the formation phase (0 – 20 Myr) which follows the core accretion paradigm, and the evolution phase (20 Myr – 10 Gyr) which analyses the thermodynamical development of each planet individually. A detailed description of the model is provided in \citeauthor{ngpps2} \citeyearpar{ngpps1,ngpps2}, with additional reviews by \cite{benz2014} and \cite{mordasini2018}.

\section{Planet engulfment in NGPPS: statistics} \label{results}

The formation and evolution of a planetary system can be a very active and chaotic process \citep{1999Natur.402..599M,2008SciAm.298e..50C}. 
The majority of planets in our sample end up being accreted by other planets, leaving less than one third remaining after 10 Gyr. Furthermore, a small fraction is either ejected from the system, disappears due to photoevaporation or falls into the star, eventually being engulfed by their host.
However, not all stars register an engulfment event. Out of 1000, 47.1\% of stars never engulfed a planet, implying that only around half of them were polluted by in-fall material from their system.

Despite there being 529 stars that engulf planetary material, signatures of these events might not be observable. A large stellar convective layer means that interior mixing processes will dilute any refractory enrichment from planetary material and over time dilute engulfment signatures \citep{2018A&A...618A.132K}.
Since the size of convective zones changes with stellar evolution, the timing of engulfment events affects the observability of the resulting signatures \citep{2020MNRAS.493.5079B,2023MNRAS.518.5465B}. Therefore, we start by analysing when do these events happen in the span of 10 Gyr and how much material falls onto the host stars.

\subsection{Timing of planet engulfment} \label{time}

\begin{figure}[H]
    \centering
    \includegraphics[scale=0.5]{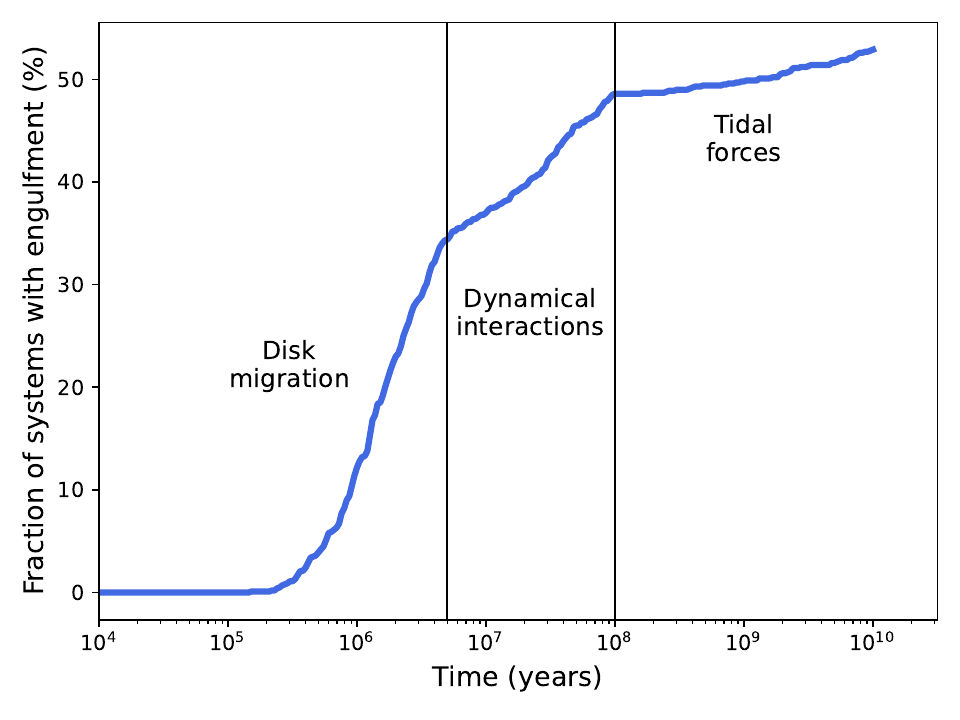}
    \caption{Percentage of systems with at least one engulfment event, for the 10 Gyr.}
    \label{events}
\end{figure}

Engulfment events can happen during the formation and evolution of planetary systems. However, they are not uniformly distributed across 10 Gyr.
Figure \ref{events} shows the cumulative distribution of the number of stars (and hence systems) with engulfment as a function of time. Although a star may engulf planets at different points in time, we account only the time of the first occurrence.
The first engulfment events happens at $\sim 1.5\times 10^5$ yr and the number of stars with at least one of these events increases with time. As shown by Figure \ref{events}, there are three main engulfment phases that reflect the physics behind the planet formation simulations: disk migration (0  $\sim$ 10 Myr), dynamical interactions (10 - 100 Myr) and tidal forces (100 Myr - 10 Gyr). 
We note we generally define $0 \sim 10$ Myr for the first phase, which lasts only until the disk dissipates, but the disk lifetime varies for each system, with the mean value being $\sim 4$ Myr.
Additionally, Table \ref{eng_fraction} reveals how the engulfment rate varies in each phase and how it affects the number of systems with engulfment.

The first transition at $5\times10^{6}$ yr marks the first change and is close to the average lifetime of the protoplanetary disks in the simulation ($\sim 4$ Myr). During this period, disk migration is the main process responsible for the engulfment of planetary material. The presence of the disk leads to planets gradually migrating closer to the star while accreting solid and gaseous material. If they migrate too close to the star, they will fall into and be engulfed by the stellar host. Over this phase, the number of systems with engulfment events increases quickly due to the effect of disk migration, leading to 370 systems with engulfment.

After the disk dissipates, this process can no longer affect the planets. However, the number of systems registering engulfment events continues to grow, albeit more slowly. Up to 100 Myr, dynamical interactions prompts collisions and scattering of planets. Due to this scattering, some planets are pushed into the star to then ultimately be engulfed by it. During this phase, 220 systems register at least one engulfment event with a total of 440 planets engulfed.

After 100 Myr, the immediate output from the model needs to be carefully considered, since dynamical interactions between planets are stopped in the model. At this stage, the model indicates there is no more accretion of either gaseous or solid material, ceasing planetary growth. The model changes processes and tidal effects are now the main mechanism affecting the system and that might lead to the star engulfing a planet. During this last stage, the number of systems with engulfment increases at the slowest rate, and we have 380 planets engulfed among 222 systems. Despite these values being similar to the ones from the previous phase, we note the difference in the time scale: these events occurred over about 10 Gyr, while the previous ones were within less than 100 Myr.
Tidal effects are stronger the more massive the planets are, since the gravitational force is stronger. 
Therefore, it is required for a system to possess a giant planet for tides to play a role, and this is not the case for all systems.

\begin{figure}
    \centering
    \includegraphics[scale=0.45]{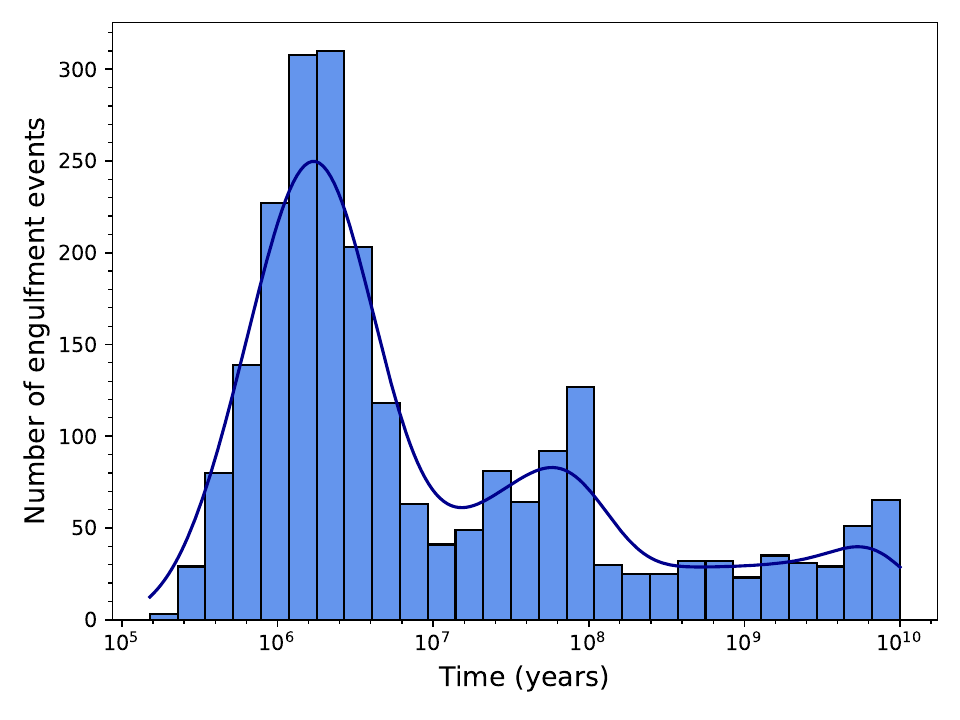}  
    \caption{Number of engulfment events at different points in time over 10 Gyr, in logarithmic scale. The solid line represents the KDE of the distribution.}
    \label{3picos}
\end{figure}

Although the number of systems with engulfment increases overtime, the frequency of these events is not always constant. In Figure \ref{3picos}, we show the distribution of the number of engulfment events as a function of time. 
The solid line represents the kernel density estimate (KDE) of the distribution.  It is possible to see how the distribution of the number of engulfment events has three peaks, namely around $\sim 2$ Myr, $\sim 100$ Myr and $\sim 8$ Gyr, and how these match to the three periods described before. 

\begin{table}[]
\caption{Fraction of planets engulfed in each phase, cumulative percentage of systems with at least one engulfment event, and number of systems with planet engulfment for each phase.}
\label{eng_fraction}
\resizebox{\columnwidth}{!}{%
\begin{tabular}{c|c|c|c|}
\cline{2-4}
 &
  \begin{tabular}[c]{@{}c@{}}Disk migration\\ (0 - 10 Myr)\end{tabular} &
  \begin{tabular}[c]{@{}c@{}}Dynamical interactions\\ (10 - 100 Myr)\end{tabular} &
  \begin{tabular}[c]{@{}c@{}}Tidal forces\\ (100 Myr - 10 Gyr)\end{tabular} \\ \hline
\multicolumn{1}{|c|}{\begin{tabular}[c]{@{}c@{}}Fraction of planets\\ engulfed (\%)\end{tabular}}       & 1.486 & 0.442 & 0.384 \\ \hline
\multicolumn{1}{|c|}{\begin{tabular}[c]{@{}c@{}}Systems with engulfment\\ (cumulative \%)\end{tabular}} & 37.0  & 48.6  & 52.9  \\ \hline
\multicolumn{1}{|c|}{\begin{tabular}[c]{@{}c@{}}Systems with \\ engulfment (per phase)\end{tabular}}    & 370   & 220   & 222   \\ \hline
\end{tabular}
}
\end{table}

\subsection{Masses of engulfed planets} \label{mass}

During the first million years planets grow in mass, and the different engulfment mechanisms are efficient for different planetary masses. As such, the distribution of the masses of the engulfed planets is not uniform across time. Moreover, the multiplicity of planet engulfment event is also an important parameter to consider as it can vary from one event to thirty. Furthermore, the mass of engulfed planets can be in range $0.01\ M_\oplus - 690\ M_\oplus$ ($\sim2.17\ M_J$).

\begin{figure}
    \centering
    \includegraphics[scale=0.45]{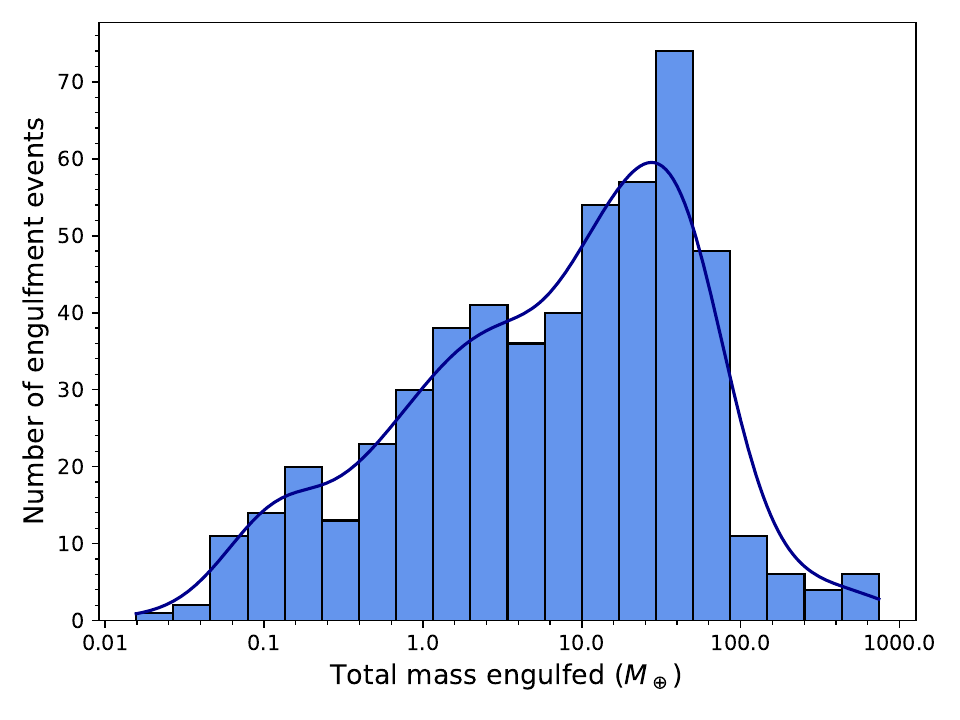}
    \caption{Total mass engulfed by the star, for each system, in logarithmic scale. The solid line represents the KDE of the distribution.}
    \label{mass_tot}
\end{figure}

Looking at the total amount of mass engulfed by each star (Figure \ref{mass_tot}), it is possible to see that there are two main  peaks. The first is around low-mass objects, with mass up to $\sim1\ M_\oplus$. The second is centred around more massive planets, with mass $M\sim 10-30\ M_\oplus$. Additionally, there is the smaller, third peak corresponding to gas giants with mass $M\sim 100-200\ M_\oplus$.

\begin{figure}
    \centering
    \includegraphics[scale=0.45]{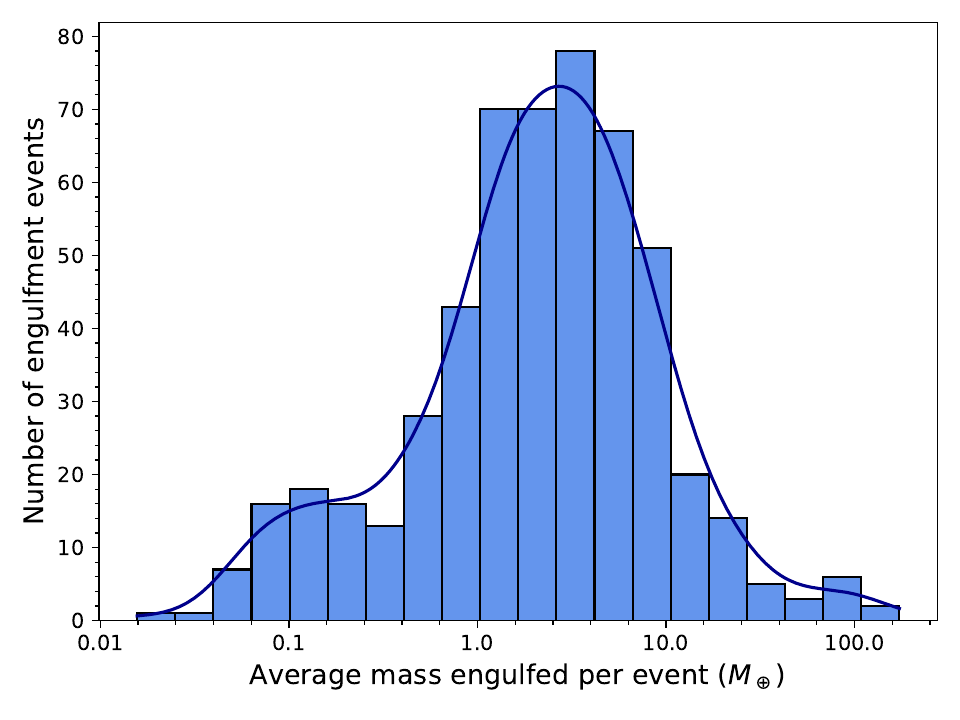}
    \caption{Average mass engulfed by the star per event, for each system. The solid line represents the KDE of the distribution.}
    \label{avg_mass_event}
\end{figure}

By accounting for the number of events and the total mass engulfed in each system, we get the distribution of the average mass engulfed per event, which can be seen in Figure \ref{avg_mass_event}. 
Analogous to Figure \ref{mass_tot}, the distribution shows two peaks, with the first centred around lower values of mass ($\sim 0.1\ M_\oplus$) and the second around higher amounts ($\sim 5\ M_\oplus$).
The wide range of masses corresponds to different type of planets engulfed.

\subsection{Planet engulfment across time and mass}

The simultaneous analysis of the mass of engulfed material and the point in time when these events happen allows to better understanding processes behind planet engulfment as a whole.

\begin{figure}
    \centering
    \includegraphics[scale=0.475]{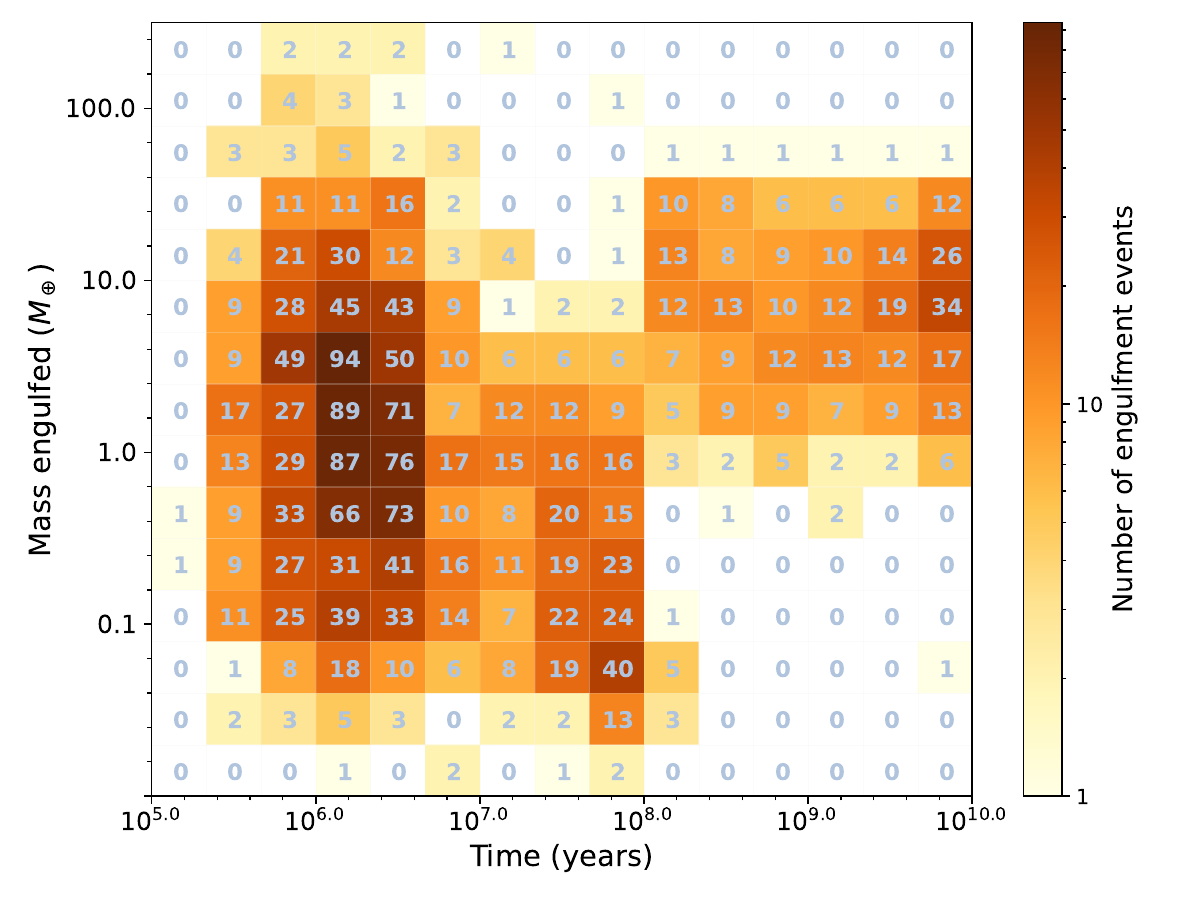}
    \caption{2D histogram representing the timing and amount of material engulfed by the central star, for 1000 systems and across 10 Gyr.}
    \label{2dhist}
\end{figure}

In Figure \ref{2dhist} we show these two quantities together colour coded by the number of engulfment events. The figure reveals the peaks observed in Figure \ref{3picos}, but now with the additional information regarding mass.
A deeper look into the data reveals how in early times (before 100 Myr), the majority of the (proto)planets engulfed were low-mass planets, which were engulfed across several events.

There are two main peaks of engulfment events during this period, once again one right up to 4 Myr, and another around 100 Myr. Throughout the first phase of engulfment, most engulfed planets have masses between M $\sim 0.6 - 5\ M_\oplus$
On the other hand, during the dynamical interactions phase, mostly planets with M $\sim 0.04 - 0.1\ M_\oplus$ are engulfed.
The second peak may be the result from the ceasing of dynamical interactions, right after 100 Myr, as explained in Section \ref{time}.

At later times (after 100 Myr), tidal effects are the ones responsible for the engulfment of material into the central star. More massive planets have a stronger impact on the stellar host, leading to stronger tidal effects. As the planet pulls the tidal bulges on the star forward, making the star spinning faster, it loses some of its angular momentum in its orbital motion around the star into the spin of the host (see \citealt{ngpps1} for a detailed description on the tidal model). 
Consequently, the planet ends up migrating closer to the star. As such, the typical mass engulfed shifts to $M\sim 10-20\ M_\oplus$.

\section{Parameters impacting planet engulfment} \label{parameters}

To better understand the processes driving infall and the resulting planetary accretion, it is crucial to identify the relevant features. 
In this section, we explore which parameters have an impact in planet engulfment.

\subsection{Parameters impacting planet engulfment: an Astrophysical approach}

\begin{figure*}[h!]
\centering
    \includegraphics[scale=0.45]{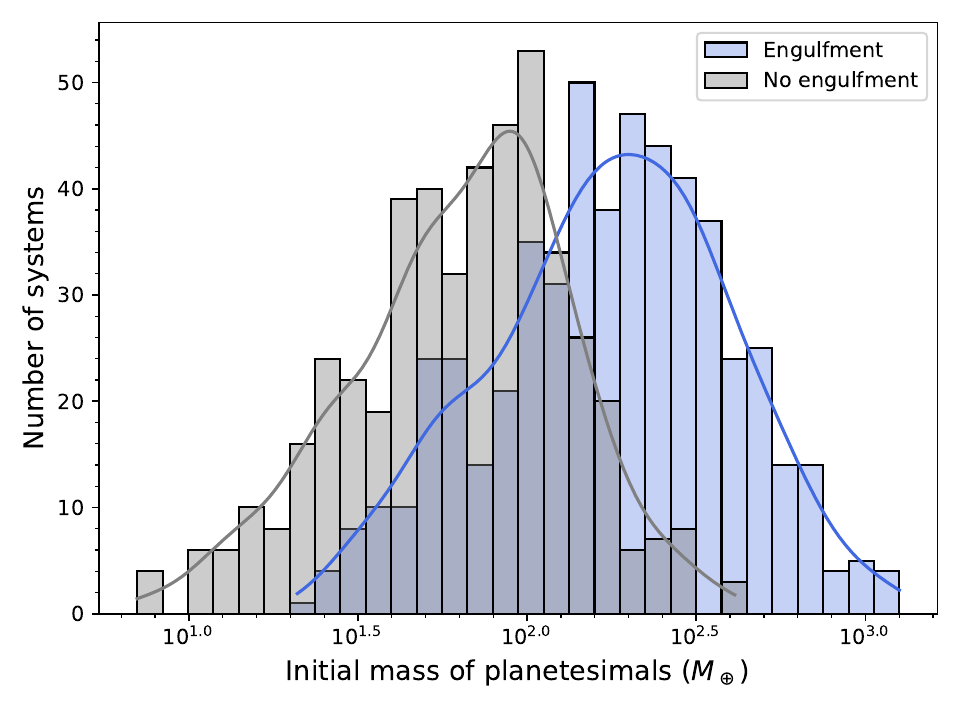}
    \includegraphics[scale=0.45]{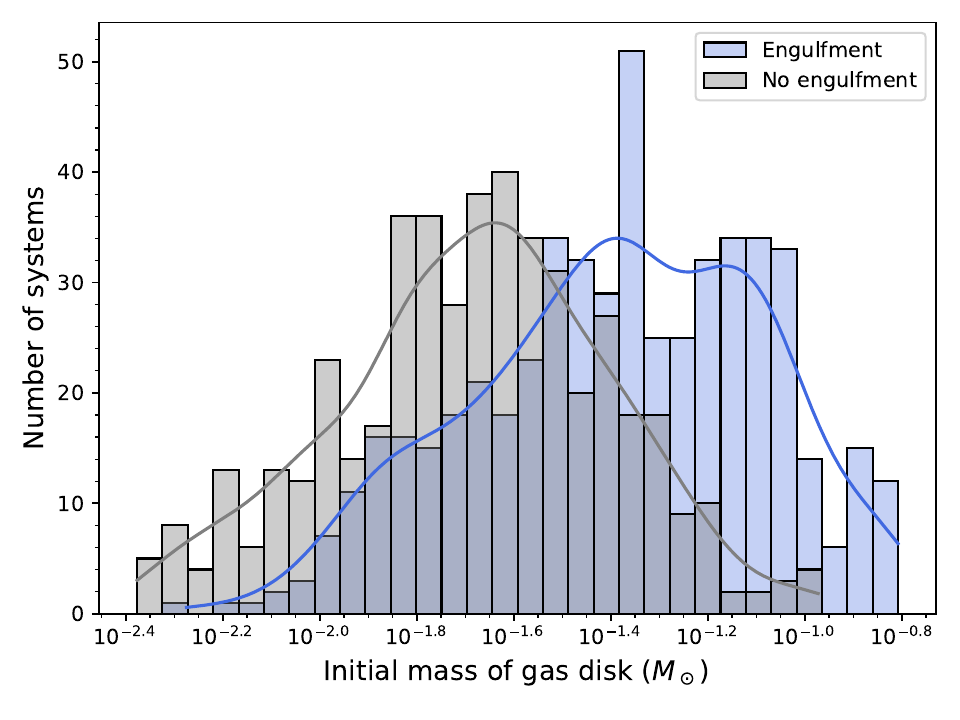}
    \includegraphics[scale=0.45]{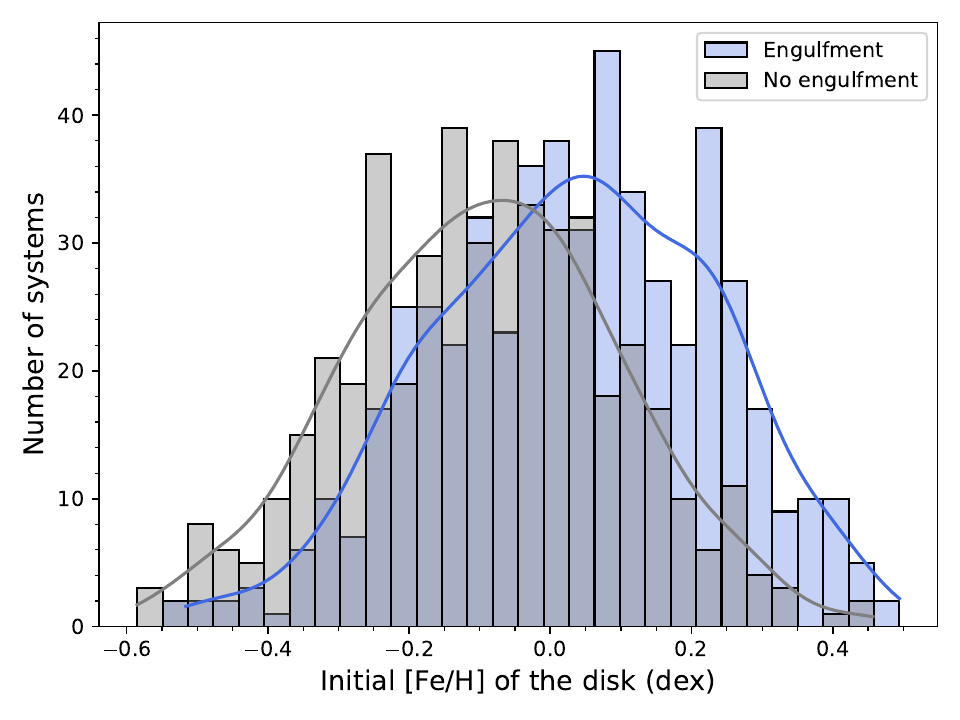}
    \includegraphics[scale=0.45]{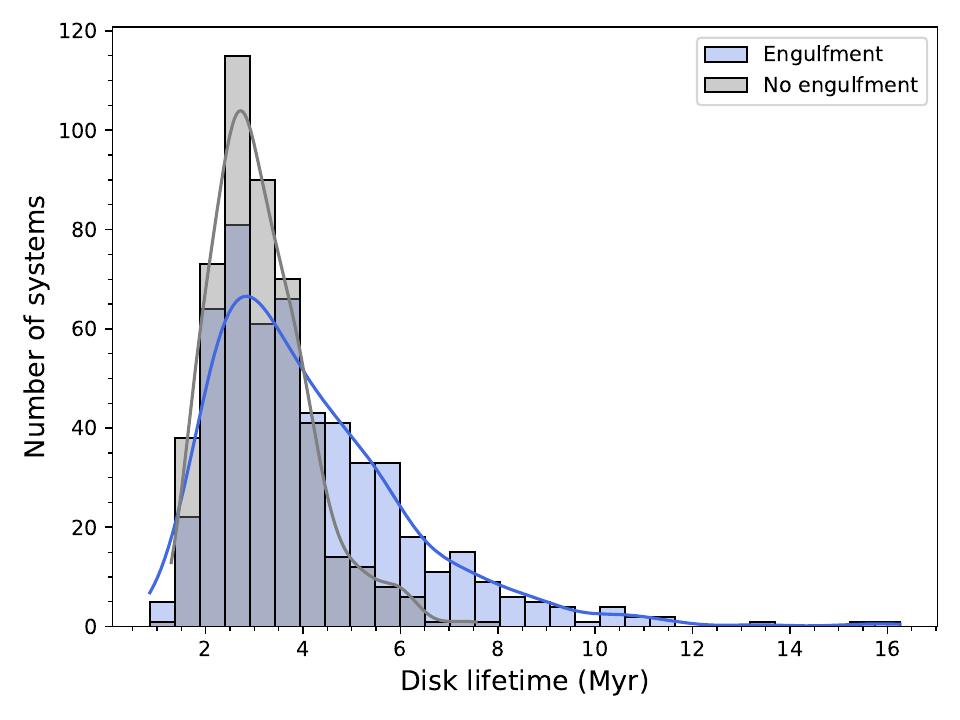}
    \caption{Initial conditions of the protoplanetary disk, for engulfing and non-engulfing systems. We show the initial mass of solids (planetesimals) in the disk (top left), and initial mass of gas disk (top right). On the bottom, we show metallicity (bottom left) and disk lifetime (bottom right).}
    \label{init}
\end{figure*}

Along with the previous analysis, the inherent properties of the protoplanetary disk also help to determine if a system is prone to undergo engulfment events or not.
Looking at the initial properties of the disk for both engulfing and non-engulfing systems can help us better understand what might affect the occurrences of these events. Figure \ref{init} shows the distribution of four parameters directly related to the initial conditions in the protoplanetary disk: initial mass of solid material (dust), initial mass of gaseous material, metallicity ([Fe/H]) and disk lifetime.
We note how for the first three parameters, there is a shift between engulfing and non-engulfing systems. The statistics of Anderson-Darling and Kolmogorov–Smirnov tests show a \textit{p-value} $< 10^{-7}$ for these three cases.

Figure \ref{init} seems to suggest that high disk mass and metallicity lead to planet engulfment.
Under such conditions, planet cores grow faster and they more quickly start accreting gas, prompting the formation of giant planets (see a detailed analysis in \citealt{ngpps2}). In turn, giants will migrate during their formation and evolution, which will generate some scatter of the planets around them, leading to some being pushed towards the star.

\begin{figure*}[h!]
    \centering
    \includegraphics[scale=0.5]{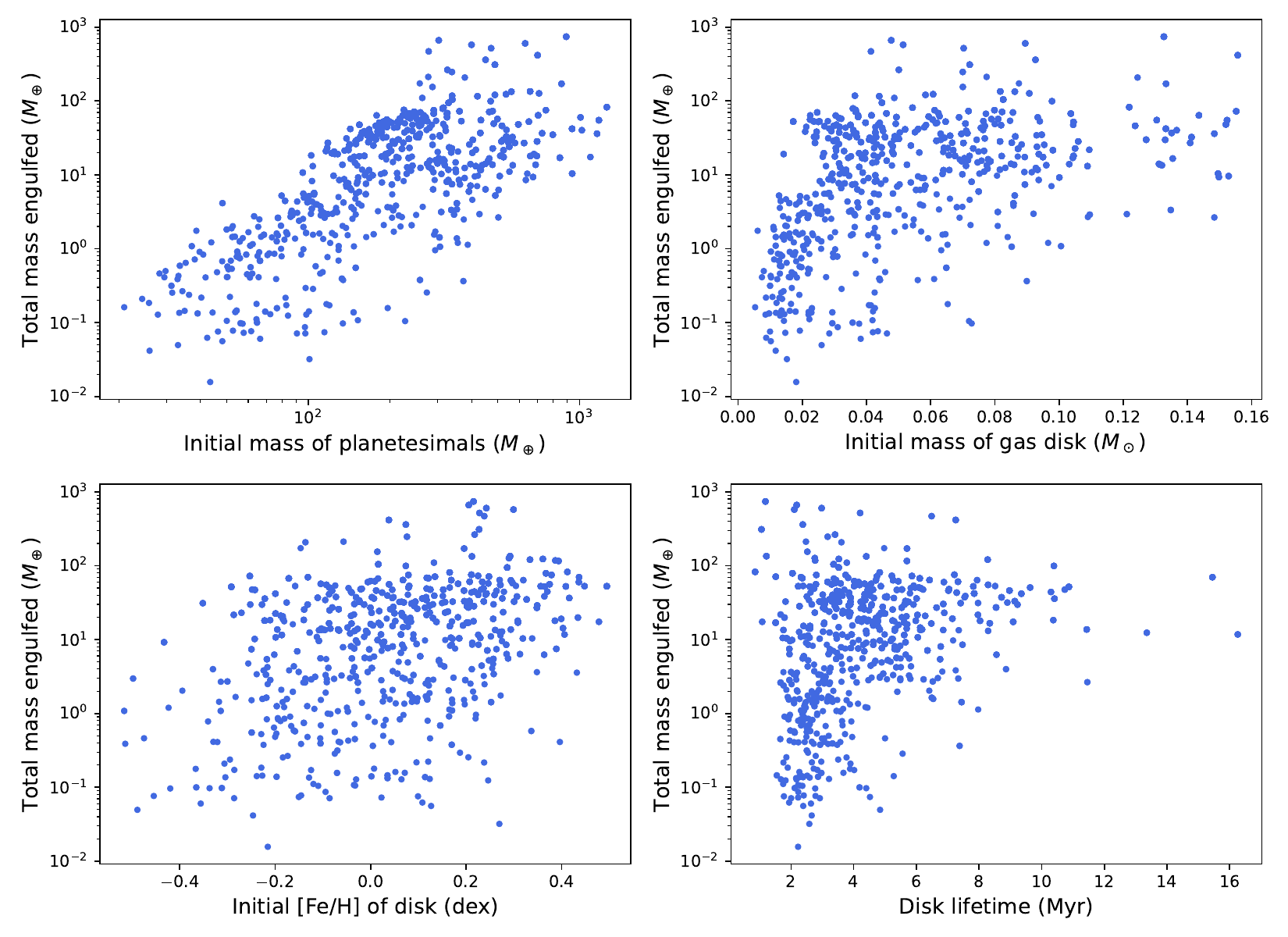}
    \caption{Total mass engulfed depending on initial conditions of the disk, for engulfing systems.}
    \label{meng}
\end{figure*}

For the systems that register at least one engulfment event, in Figure \ref{meng} we look at the total amount of mass engulfed in the system, depending on four initial parameters previously mentioned that reveal a difference between populations.
There is a clear correlation with initial mass, as systems with high initial mass tend to engulf a higher amount (since there is more material available). In addition, the more metallic a system is, the more mass it tends to engulf and though not as obvious, there is also a correlation between these two parameters (Spearman coefficient of 0.33).
The dependence we found for these parameters seem to support the idea that systems with conditions more favourable to the formation of giants are more likely to experience planet engulfment.

Although the presence of giant planets can lead to planet engulfment, it is important to notice the direction of the causality relation between initial conditions and presence of giant planets is not clear.
The presence of giant planets in the system are not necessarily the main reason behind the engulfment of planets, as the formation of giant planets also depends on the same aforementioned initial conditions of the disk.
Furthermore, although systems with more massive and more metallic disks are more prone to form giants and undergo planet engulfment, there exist cases of systems that also register engulfment events but do not form giant planets. As such, giant planets cannot be the sole reason behind planet engulfment.

\subsection{Parameters impacting planet engulfment: a Machine Learning approach}

On top of the initial conditions of the planet-forming disks, the NGPPS database contains over 140 columns of information about the characteristics of planetary embryos at a given time snapshot.
In order to understand which parameters are more important for a correct prediction of engulfment in planetary systems, we used several Machine Learning (ML) supervised classification models to analyse the data. Our aim was to identify the astrophysical parameters that are most relevant when predicting whether engulfment will happen in a system. For this, we utilized the XGBClassifier\footnote{\url{https://docs.getml.com/latest/api/getml.predictors.XGBoostClassifier.html}} from XGBoost \citep{XGBoost}, the RandomForestClassifier from Random Forest \citep{RF}, and the SVC from Support Vector Machines \citep{SVM} as implemented in scikit-learn\footnote{\url{https://scikit-learn.org}}. We also attempted to use Logistic Regression \citep{LogReg} from scikit-learn; however, due to the non-linearity of the dependencies, this model did not perform well.

We performed a standard analysis of the data, including exploratory data analysis, normalization,  addressing multicollinearity using the variance inflation factor. We split the data into training and testing sets, and optimised the hyperparameters of the models to achieve the best performance for the classification of engulfed and non-engulfed systems. Through these models, we achieved an accuracy of approximately 88\% in predicting whether a system would experience an engulfment event. The five critical parameters influencing the models to attain this level of accuracy were the maximum planetary mass within the system at 10 Gyr, the minimum orbital distance of planets at 10 Gyr, the disk lifetime, initial metallicity, and the initial mass of planetesimals in the disk.

\begin{figure*}[]
    \centering
    \includegraphics[scale=0.52]{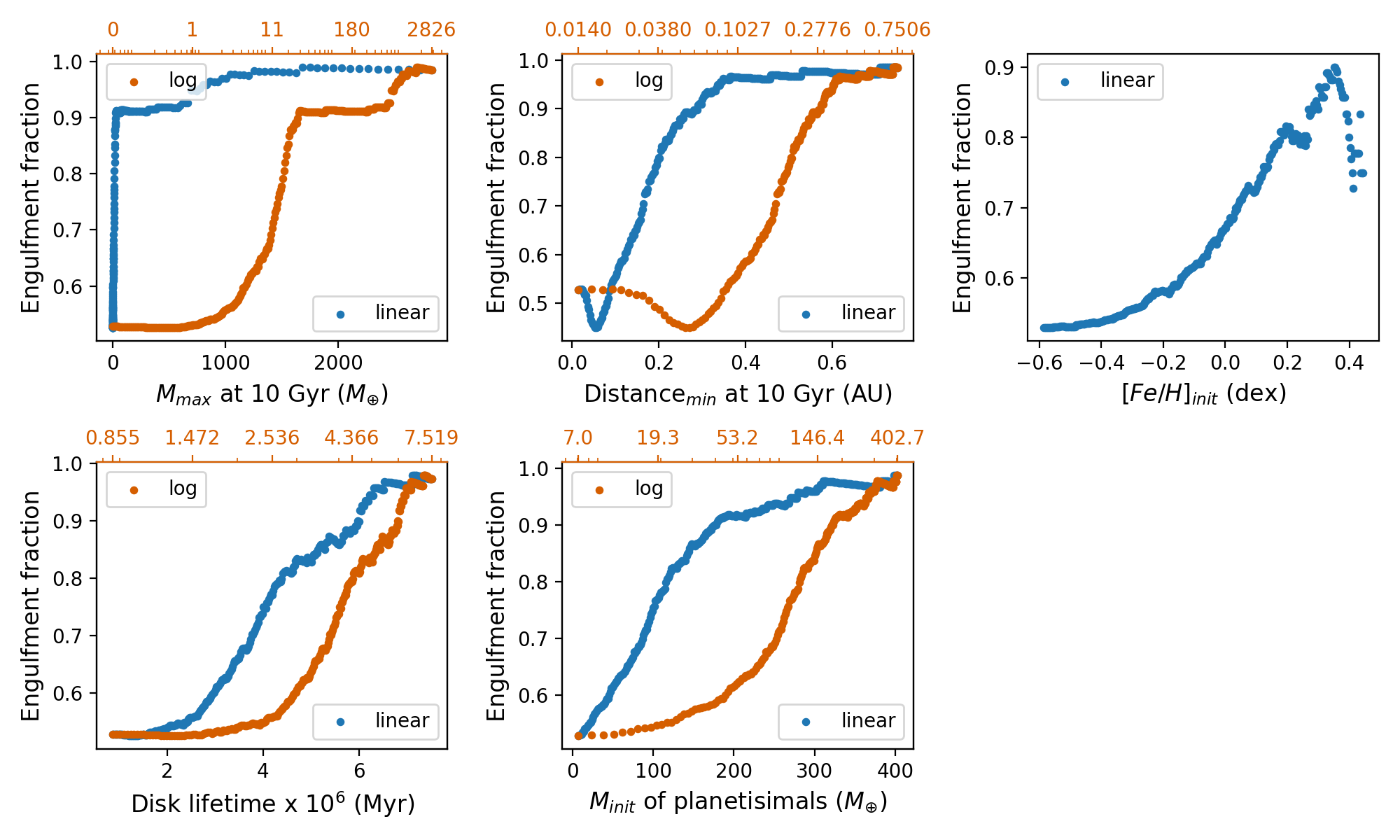}
    \caption{Fraction of systems with engulfment depending on the maximum planetary mass within the system at 10 Gyr, the minimum orbital distance of planets at 10 Gyr, the disk lifetime, initial metallicity, and the initial mass of planetesimals in the disk, in linear scale (blue) and logarithm scale (red).}
    \label{ml}
\end{figure*}

In Figure \ref{ml}, we illustrate how the fraction of systems with engulfment is contingent on the aforementioned five parameters. The figure shows that these dependencies are not necessarily linear, and for some parameters, there are thresholds beyond which the majority of systems with these properties experience engulfment. For instance, more than 90\% of systems with a massive planet $(M > 30\ M_\oplus)$ and minimum orbital distances of $a>0.3\ \mathrm{AU}$ undergo engulfment during their evolution. Moreover, there are thresholds above which all the systems experience engulfment. Specifically, all 64 systems with a maximum mass of a planet of $M > 2900\ M_{\oplus}$ ($M>9\ M_J$) experience engulfment. The figure also shows that systems that are overall more metal-rich and have more massive and longer-lived disks are more likely to undergo planet engulfment, as previously concluded.

In the plots of the minimum distance and initial [Fe/H] from Figure \ref{ml}, we observe a decline in the overall increase of the engulfment fraction for $a_{min,10\ Gyr}<0.05$ AU and [Fe/H]$_{init}>0.3$ dex, respectively. 
For the first case, this drop occurs because there are many engulfing systems with planets orbiting at distances less than 0.05 AU (at 10 Gyr), resulting in a cluster of systems within the range $0< a_{min,10gyr} < 0.05$ AU on the $ a_{min,10gyr}$ distribution. 
%
Thus, when we consider higher values up to 0.05 AU for the minimum distance, we are excluding these very close-by systems. This leads to a sharp decrease of the number of engulfing systems compared to the significantly larger number of non-engulfing systems, resulting in a reduced fraction. Beyond 0.05 AU, the systems are more evenly spread.
Analogously, there are very few systems with [Fe/H] $>$ 0.3 dex (less than 5\%). When we increase the value of the initial [Fe/H] above 0.3 dex, the total number of systems with planet engulfment and such value of [Fe/H] decreases.

\section{Composition of engulfed and remaining planets} \label{variation composition}

During its lifetime, a star can engulf planets with a diverse range of masses and compositions, which could then affect the stellar atmosphere composition.

To better understand the chemical patterns of the engulfed planetary material compared to that of the planets that remain on stable orbits, we select 9 cases to analyse:
engulfment of planets with mass $M = 2$, $5$ and $10\ M_\oplus$, at times around $8$ Myr, around $60$ Myr and around $4.5$ Gyr.
The first period should not be too short in order to avoid for the presence of the disk and the large stellar convective layer, hence we chose 8 Myr. The second period was selected to be around 60 Myr. Finally, we decided on 4.5 Gyr for the last one as this is around the solar age, and our results show the impact on stars is not very different at the late stage of evolution ($\gtrsim 1$ Gyr).

Since the Bern model does not include a continuous time distribution but provides instead snapshots of different points in time, and there are not enough events happening at a given time, we consider the aforementioned times within a certain interval: $8\pm 2$ Myr, $60 \pm 40$ Myr and $4.5 \pm 1$ Gyr.
For the previously mentioned masses, in the period between 10 and 100 Myr, the number of engulfment events is small, and thus the model does not provide enough information to estimate representative average values. As such, we decided to extend the period around 60 Myr to account for more events.
Likewise, in the simulated data, the planets' mass is not exactly $2, 5$ or $10\ M_\oplus$. As such, we consider a range of masses centred on these values, and in a way that the intervals do not overlap: $2 \pm 1.5\ M_\oplus$, $5 \pm 1.5\ M_\oplus$ and $10 \pm 3.5\ M_\oplus$.
The size of the intervals was adjusted in order to guarantee a minimum number of three events in each case, for the different periods.

For each instance, we analyse the chemical composition of the planetary material. The internal structure model from Generation III Bern model assumes that planets have a spherically symmetric interior with an onion-like structure composed by an iron core, a silicate mantle and (depending on a planet's accretion history) a water ice layer and a gaseous envelope made of pure H/He.
In this work, we will refer to the gaseous layer surrounding the planet as the 'atmosphere', and the remaining layers of the planet, not including the atmosphere, as the 'core'.

The Generation III Bern model provides information on the ice mass fraction of the core, $f_{ice}$, and the iron mass fraction in the core's refractory component, $f_{iron}$. We compute the corresponding mass fraction of ice, iron and silicates of each planet from the core mass, $M_{core}$:

\begin{align}
   & M_{ice} = M_{core} \times f_{ice} \nonumber \\
   & M_{iron} = (M_{core} - M_{ice})  \times f_{iron} \\
   & M_{silicates} = M_{core} - M_{ice} - M_{iron} \nonumber
\end{align}

Since the simulations contain only information on the amount of iron, silicates and ices without discriminating specific elements, we take a simple approach regarding the chemical composition.

We start by assuming the ices are composed only by water and disregard highly volatile molecules such as $\mathrm{CH_4}$, $\mathrm{CO}$ and other carbon-bearing species. This choice is based on results from \cite{ices} where $\mathrm{H_2 O}$ is identified as the dominant chemical species in ices and composing $\sim 60\%$ of its mass.
For the silicates, we use the Earth composition as reference and following \cite{earth_comp}, we divide the silicates into 80\% $\mathrm{MgSiO_3}$ (perovskite, a magnesium silicate) and 20\% $\mathrm{Mg_{2}SiO_4}$ (olivine).
For the atmosphere, we assumed that 76\% of the atmosphere mass ($M_{atm}$) corresponds to hydrogen and the remaining 24\% to helium, considering the fixed value of helium mass fraction in the simulations is $\mathrm{Y=0.24}$.

With information of these minerals, together with iron, water and an atmosphere made of H/He (if present), we extracted the mass fraction of H, He, O, Mg, Si and Fe. 
Through stoichiometric relations and the molecular mass of the different species, we compute the mass of the different elements using:
\begin{equation}
    m_X = \alpha.n_Y \times \mathcal{M}_X
\end{equation}
where $X$ represents the element we want to determine the mass of, $\alpha$ its stoichiometric coefficient, $\mathcal{M}$ the molecular mass, $n$ the moles number  and $Y$ the molecule it is part of.

Some species like He and Fe are straightforward to calculate, since they come directly from $M_{atm}$ and $M_{iron}$. However, H, O, Mg and Si are present in more than one molecules and therefore all contributions need to be accounted for:


\begin{align}
    M_{H} &=  0.76 \times M_{atmo} + 2n_{H_2O} \times \mathcal{M}_{H} \nonumber \\ 
    M_{He} &=  0.24 \times M_{atmo} \nonumber \\ 
    M_{O} &=  (3n_{MgSiO_3} + 4n_{Mg_2 SiO_4} + n_{H_2O}) \times \mathcal{M}_{O} \\ 
    M_{Mg} &=  (n_{MgSiO_3} + 2n_{Mg_2 SiO_4}) \times \mathcal{M}_{Mg} \nonumber \\ 
    M_{Si} &=  (n_{MgSiO_3} + n_{Mg_2 SiO_4}) \times \mathcal{M}_{Si} \nonumber \\
    M_{Fe} &=  M_{iron} \nonumber
\end{align}

In Table \ref{mass_frac} we present the results of the calculated the mass fraction of all aforementioned elements for the engulfed planets and for those remaining in the system after 10 Gyr.

We notice that, in general, the engulfed planets have a higher amount of Mg, Si and Fe than the planets remaining in the system. This could be due to the fact that if these engulfed planets formed inside the ice line, closer to the star where condensation temperatures are higher, their composition would be richer in refractory elements like Mg, Si and Fe (see e.g. \citealt{2003Lodders,WANG2019287}). Being at a shorter orbital distance from the star, they would be more susceptible to be pushed into the host star and eventually engulfed by it.

On the other hand, the planets that remain in the systems are generally richer in oxygen. Following our simple approach in composition, this element comes not only from the olivine and perovskite, but also from the molecule of water. If these planets have lower abundances of Mg, Si and Fe, but higher amount of O, this points to a higher content of water in these planets, meaning they could have formed farther away from the host star, contrary to the engulfed planets. In the surviving planets, the amount of hydrogen compared to helium is greater than what would come exclusively from the atmosphere (76\% H, 24\% He). This means the hydrogen amount also comes from water molecules, which supports the hypothesis that these planets formed at longer orbital distance from the star, and thus are less likely to be engulfed by the host star.


\begin{table}[]
\caption{Average percentage mass fraction of elements in engulfed planets (top) and remaining planets (bottom).}
\label{mass_frac}
\resizebox{\columnwidth}{!}{%
\begin{tabular}{|c|cccccc|c|c|}
\hline
\multirow{2}{*}{\textbf{Time}} & \multicolumn{6}{c|}{\textbf{Mass fraction (\%)}} & \multirow{2}{*}{\textbf{Nr. events}} & \multirow{2}{*}{\textbf{Engulfed mass}} \\ \cline{2-7}
                 & \textbf{H} & \textbf{He} & \textbf{O} & \textbf{Mg} & \textbf{Si} & \textbf{Fe} &    &                                  \\ \hline
$8 \pm 2$ Myr    & 0          & 0           & 32         & 18          & 18          & 32          & 24 & \multirow{3}{*}{2 $M_{\oplus}$}  \\
$60 \pm 40$ Myr & 0          & 0           & 33         & 18          & 18          & 30          & 64 &                                  \\
$4.5 \pm 1$ Gyr  & 0          & 0           & 32         & 18          & 18          & 32          & 10 &                                  \\ \hline
$8 \pm 2$ Myr    & 4          & 1           & 44         & 15          & 15          & 21          & 4  & \multirow{3}{*}{5 $M_{\oplus}$}  \\
$60 \pm 40$ Myr & 0          & 0           & 32         & 18          & 18          & 32          & 9  &                                  \\
$4.5 \pm 1$ Gyr  & 0          & 0           & 34         & 19          & 19          & 28          & 14 &                                  \\ \hline
$8 \pm 2$ Myr    & 1          & 0           & 41         & 17          & 17          & 23          & 4  & \multirow{3}{*}{10 $M_{\oplus}$} \\
$60 \pm 40$ Myr & 0          & 0           & 33         & 18          & 18          & 30          & 4  &                                  \\
$4.5 \pm 1$ Gyr  & 3          & 0           & 51         & 14          & 14          & 17          & 8  &                                  \\ \hline
\end{tabular}%
}

\vspace{0.2cm}

\resizebox{\columnwidth}{!}{%
\begin{tabular}{|c|cccccc|c|c|}
\hline
\multirow{2}{*}{\textbf{Time}} & \multicolumn{6}{c|}{\textbf{Mass fraction (\%)}} & \multirow{2}{*}{\textbf{Nr. planets remaining}} & \multirow{2}{*}{\textbf{$M_{planet}$}} \\ \cline{2-7}
                 & \textbf{H} & \textbf{He} & \textbf{O} & \textbf{Mg} & \textbf{Si} & \textbf{Fe} &     &                                  \\ \hline
$8 \pm 2$ Myr    & 4          & 0           & 54         & 13          & 13          & 17          & 419 & \multirow{3}{*}{2 $M_{\oplus}$}  \\
$60 \pm 40$ Myr & 5          & 0           & 61         & 10          & 10          & 13          & 602 &                                  \\
$4.5 \pm 1$ Gyr  & 13         & 3           & 55         & 9           & 9           & 11          & 7   &                                  \\ \hline
$8 \pm 2$ Myr    & 4          & 0           & 56         & 12          & 12          & 15          & 291 & \multirow{3}{*}{5 $M_{\oplus}$}  \\
$60 \pm 40$ Myr & 5          & 0           & 56         & 12          & 12          & 15          & 393 &                                  \\
$4.5 \pm 1$ Gyr  & 5          & 0           & 58         & 11          & 11          & 14          & 6   &                                  \\ \hline
$8 \pm 2$ Myr    & 4          & 0           & 56         & 12          & 12          & 15          & 110 & \multirow{3}{*}{10 $M_{\oplus}$} \\
$60 \pm 40$ Myr & 4          & 0           & 54         & 13          & 13          & 16          & 263 &                                  \\
$4.5 \pm 1$ Gyr  & 7          & 2           & 44         & 15          & 15          & 18          & 1   &                                  \\ \hline
\end{tabular}%
}
\end{table}


\section{Impact of planet engulfment on stellar composition} \label{morgan}

In order to assess the effect of planet engulfment on the surface composition of the host star, stellar models are computed with the Cesam2k20\footnote{\url{https://www.ias.u-psud.fr/cesam2k20/home.html}} (Code d'Evolution Stellaire Adaptatif et Modulaire) stellar evolution code \citep{morel08,marques13,deal18}. The models are computed with $X_0=0.720$, $Y_0=0.264$, and $Z_0=0.016$ (solar calibrated) and an initial mixture of metals following the solar abundances determined by \cite{asplund21}, the OPAL equation of states \citep{rogers02} and opacity tables \citep{iglesias96} complemented at low temperature by the table of \cite{ferguson05}.
The nuclear reaction rates are taken from the NACRE (Nuclear Astrophysics Compilation of REactions rates) compilation \citep{angulo99}, except for the $^{14}\mathrm{N}(p,\gamma)^{15}\mathrm{O}$ reaction, for which we adopted the LUNA (Laboratory for Underground Nuclear Astrophysics) rate \citep{imbriani04}. Convection is computed following the CGM formalism \citep{canuto96} with a solar calibrated parameter $\alpha_\mathrm{CGM}=9183$. We used the \cite{krisw66} $T(\tau)$ relation for the atmosphere.
For the transport of chemical elements, we included atomic diffusion following the \cite{michaud93} formalism and thermohaline convection is included with the \cite{brown13} prescription. When an additional transport is included, a turbulent diffusion coefficient is added with the expression
\begin{equation}
    D_\mathrm{turb}=\omega D(\mathrm{He})_0\left(\frac{\rho_0}{\rho}\right)^n
\end{equation}
\noindent where $\omega=400$ and $n=3$ are constants, $\rho$ is the density, and $D(\mathrm{He})_0$ and $\rho_0$ are the diffusion coefficient of helium and density at the reference depth defined with the temperature $T_0$, respectively \citep[see][]{richer00}. The reference depth is calibrated to produce a transport to obtain the lithium surface abundance of the Sun for a solar model. The calibrated value is $\log(T_0)=6.44$.
When accretion is taken into account, the matter is instantaneously diluted in the convective envelope with an accretion rate of $M_\mathrm{accr}/1\mathrm{Myr}$ during $1$~Myr after the start of the accretion. Any additional dilution directly comes from the transport in the radiative zone.

\begin{figure}[H]
    \centering
    \includegraphics[width=\columnwidth]{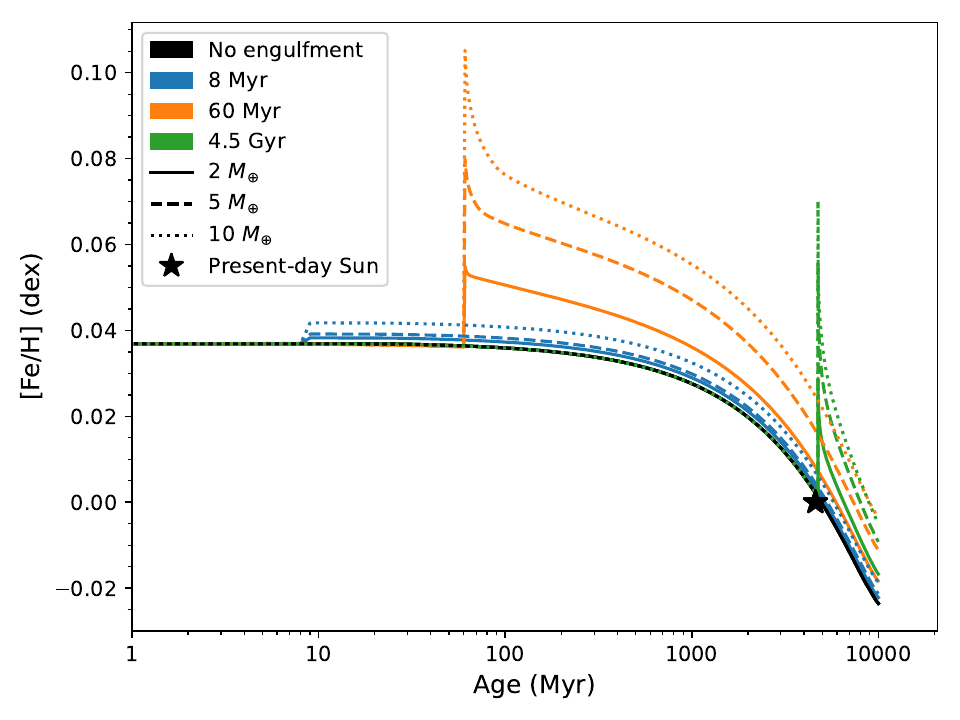}
    \caption{Variation of iron abundance on the stellar surface due to an engulfment event occurring at 8 Myr (blue), 60 Myr (orange) or 4.5 Gyr (green). Included are the cases for an engulfed planet with mass of $2\ M_{\oplus}$ (solid line), $5\ M_{\oplus}$ (dashed line) or $10\ M_{\oplus}$ (dotted line) The black solid line models the evolution of the solar composition.}
    \label{output}
\end{figure}

Through these stellar models, it is possible to determine the impact of planet engulfment on the surface composition of the host star.
The composition of the planetary material is estimated from the simulations, as described in Section \ref{variation composition}. 
We apply the aforementioned 9 cases to the stellar models and check how a planet with a given mass engulfed by its host at a given time changes the stellar composition.
The results of the variation of [Fe/H] due to planetary engulfment can be found in Figure \ref{output}. The black solid line found in the plot corresponds to the evolution of the composition for the solar case. 

According to SP21, in our analysis we set $\sigma_{[Fe/H]} = 0.02$ dex as the minimum value for a variation in [Fe/H] to be measurable. On a more conservative approach, we also consider a threshold of 0.06 dex for the [Fe/H] variation to be observable. This number reflects the middle value of the [Fe/H] variation for the binaries classified by SP21 as chemically anomalous, i.e. the median of the data set.

The timing and amount of engulfed material pertain to the 9 cases mentioned in Section \ref{variation composition}. Regardless of the mass of the engulfed planet, events happening around 8 Myr do not have an observable impact in the stellar atmosphere's composition, as the biggest variation corresponds to an enhancement of $< 0.01$ dex, for the case of $M\sim 10 M_{\oplus}$.

As shown in Table \ref{visible}, the engulfment events that occur around 60 Myr can generate a [Fe/H] variation of 0.02 dex (case of $2\ M_{\oplus}$), 0.04 dex (case of $5\ M_{\oplus}$) and up to 0.07 dex (case of $10\ M_{\oplus}$) immediately after the accretion of the planetary material. If we consider the criteria from SP21, in all cases would be possible to detect traces of planet engulfment. However, if we assume instead $\sigma_{[Fe/H]}=0.06$ dex, only an engulfment of a planet with $10\ M_\oplus$ would provide an amount of material large enough to induce an observable change in the stellar surface composition. 
Nevertheless, as the star evolves and its internal processes dilute the additional material, this chemical imprint reduces respectively to < 0.01, 0.02 and 0.03 dex after 1 Gyr, leading to the chemical imprint of a $2\ M_\oplus$ engulfed planet no longer being detectable by SP21's criteria.
%
However, 3 Gyr after the event only traces consequent of the engulfment of a 10 $M_\oplus$ remain detectable.
On the other hand, if we assume $\sigma_{[Fe/H]}=0.06$ dex then no chemical imprint on the stellar surface is large enough to be measured 1 Gyr following the occurrence.

If we have an engulfment event of the same three cases at 4.5 Gyr, the induced signature on [Fe/H] variation is 0.03, 0.06 and 0.07 dex for the case of engulfment of a $2\ M_{\oplus}$, $5\ M_{\oplus}$ and $10\ M_{\oplus}$ planet, respectively, all detectable by SP21's criteria. Looking at 1 Gyr and 3 Gyr after the event, the internal processes of the star dilute the chemical imprint on the stellar surface and the signatures for the $2\ M_{\oplus}$ and $5\ M_\oplus$ cases are no longer observable. Moreover, if we consider $\sigma_{[Fe/H]}=0.06$ dex, no variation of the iron abundance remains detectable for these scenarios.

Though the findings presented correspond only to the variation in the iron abundance, the [O/H], [Mg/H] and [Si/H] variations follow the same behaviour, hence leading to similar conclusions.

\begin{table}[]
\centering
\caption{Variation in [Fe/H] caused by engulfment of 2, 5 and 10 $M_\oplus$ at 60 Myr and 4.5 Gyr, and evolution of the chemical signature after 1 Gyr and 2 Gyr.}
\label{visible}
\resizebox{0.9\columnwidth}{!}{%
\begin{tabular}{|c|c|c|cc|}
\hline
\multirow{2}{*}{\textbf{Time}} & \multirow{2}{*}{$M_{planet}\ (M_\oplus)$} & \multirow{2}{*}{$\Delta [Fe/H]$} & \multicolumn{2}{c|}{\textbf{$\sigma_{[Fe/H]}$ (dex)}} \\ \cline{4-5} 
                                 &    &                 & \multicolumn{1}{c|}{\textbf{0.02}} & \textbf{0.06} \\ \hline
\multirow{3}{*}{60 Myr}          & 2  & 0.02            & yes                       & no   \\
                                 & 5  & 0.04            & yes                       & no   \\
                                 & 10 & 0.07            & yes                       & yes  \\ \hline
\multirow{3}{*}{60 Myr + 1 Gyr}  & 2  & \textless{}0.01 & no                        & no   \\
                                 & 5  & 0.02            & yes                       & no   \\
                                 & 10 & 0.03            & yes                       & no   \\ \hline
\multirow{3}{*}{60 Myr + 3 Gyr}& 2  & \textless{}0.01 & no                        & no   \\
                                 & 5  & 0.01& no& no   \\
                                 & 10 & 0.02& yes                       & no   \\ \hline
\multirow{3}{*}{4.5 Gyr}         & 2  & 0.03            & yes                       & no   \\
                                 & 5  & 0.06            & yes                       & yes  \\
                                 & 10 & 0.07            & yes                       & yes  \\ \hline
\multirow{3}{*}{4.5 Gyr + 1 Gyr} & 2  & 0.01            & no                        & no   \\
                                 & 5  & 0.02            & yes                       & no   \\
                                 & 10 & 0.03            & yes                       & no   \\ \hline
\multirow{3}{*}{4.5 Gyr + 3 Gyr}& 2  & <0.01& no                        & no   \\
                                 & 5  & <0.02& no& no   \\
                                 & 10 & 0.02            & yes                       & no   \\ \hline
\end{tabular}%
}
\tablefoot{Considering a threshold of  $\sigma_{[Fe/H]} = 0.02$ dex or a more conservative threshold of  $\sigma_{[Fe/H]} = 0.06$ dex affects which events would be detectable.}
\end{table}


\section{Discussion} \label{discussion}

Several studies have addressed the subject of planet engulfment. As previously mentioned, in their work SP21 report an engulfment rate of $\sim 27\%$ for Sun-like stars.
In their analysis, they account for the presence of a convective layer in the star and note that the engulfment events must have happened after the dissipation of the protoplanetary disk, otherwise the thick stellar convective zone \citep{2018A&A...618A.132K} would dilute the accreted material without producing any substantial variation in the atmospheric chemical composition \citep{2015A&A...582L...6S}.

Nonetheless, the authors do not mention the exact moment when these occurrences should happen in order to leave a chemical imprint in the host's composition. Moreover, they do not discuss any specific stellar structure considered, despite being a necessary part to consider since, as mentioned in Section \ref{morgan}, processes like atomic diffusion, thermohaline mixing and other transport processes can affect the lifetime of the chemical patterns induced by the engulfment of planetary material.

In this work, we account for both aspects when we evaluate the impact of planet engulfment in the stellar surface composition with the stellar models computed with Cesam2k20. 
As  explained in Section \ref{morgan}, accretion of material onto the stellar surface at 8 Myr will not induce a visible variation on the [Fe/H] abundance due to the size of the stellar convective layer at this early stage \citep{2018A&A...618A.132K}. However, as the star evolves the convective zone will gradually shrink \citep{Stahler_Palla_2004}, thus the signature on the star's composition caused by planet engulfment would be increasingly greater at later ages.

In this line of reasoning, in Table \ref{frac_systems} we calculate the fraction of systems that have registered an engulfment event of a planet with $M> 2 M_\oplus, M > 5 M_\oplus$ and $M > 10 M_\oplus$ after a given age, considering the $\sigma = 0.02$ dex threshold. When we consider an event at any point after 60 Myr, $\sim 20\%$ of the systems engulf a $M>2\ M_\oplus$ planet. However, we show in Table \ref{visible} that 3 Gyr after the event, 
only the variation in [Fe/H] caused by the accretion of more massive planets ($M>10\ M_\oplus$) remains detectable.
As thus, when trying to observe the resulting signature a few Gyr after, only $\sim 11\%$ of the stars would still have traces of planet engulfment.

The more this age threshold is moved towards later ages, the less systems exist that undergo planet engulfment in these conditions since we are shortening the time interval. Furthermore, as we consider engulfment of more massive planets, the fraction of systems will also decrease, as massive planets are less common than low-mass planets \citep[see e.g.][]{Fressin_Torres_Charbonneau_Bryson_Christiansen_Dressing_Jenkins_Walkowicz_Batalha_2013,Rowan_Meschiari_Laughlin_Vogt_Butler_Burt_Wang_Holden_Hanson_Arriagada_et_al._2016,2021ApJS..255...14F}. 


To estimate the engulfment rate liable to be observed in the present day, we need to account not only for stellar age, but also the lifetime of the signal resulting from planet engulfment.
Analysing Table \ref{visible}, we find that traces from the engulfment of massive planets ($M>10\ M_\oplus$) remain detectable for several Gyr, even if the occurrence happens as early as 60 Myr. Looking at Table \ref{frac_systems}, this corresponds to $10.5\%$.
On the other hand, the lifetime of an observable signature from the engulfment of a $5\ M_\oplus$ is $\sim 2$ Gyr. 
Since the majority of stars in the solar neighbourhood are older than 3 Gyr \cite[e.g.][]{2013A&A...560A.109H, 2018MNRAS.477.2966L,2023A&A...678A..39G},  we should also include events that occur after 1 Gyr for planets with masses between $5 - 10\ M_\oplus$ (as planets more massive than $10\ M_\oplus$ are already contemplated in $10.5\%$ of the systems). The consequent chemical imprint can visible for a few Gyr, corresponding to an additional $5.6\%$ of the systems. 
In this optimistic scenario, we estimate that the engulfment rate leading to a measurable change in stellar composition is smaller than $20\%$.

SP21 report an engulfment rate that could be comparable with our results if we consider engulfment of planets with mass $M > 2\  M_\oplus$ happening after 60 Myr, with observations made  immediately after the event (Table \ref{frac_systems}). However, as shown in Table \ref{visible}, the engulfment of lower-mass planets causes an abundance difference of less than 0.03 dex on [Fe/H], which is no longer detectable 1 Gyr after the occurrence.
If we instead assume a minimum detectable variation of 0.06 dex in [Fe/H], we would be referring to the engulfment of more massive planets ($M> 10\ M_\oplus$), which comprises 11\% of the systems, less than half of the reported rate. This analysis was done assuming the more conservative scenario in which engulfment events can happen at any point after 60 Myr, with stars observed immediately following the occurrence.
This implies that some stars must still be young for the chemical imprint of the event to remain visible and not yet diluted by the stellar internal processes. In fact, Table \ref{frac_systems} and Figure \ref{2dhist} show that a large fraction of the engulfment events happen during the first 100 Myr. Nevertheless,  the stars in the solar neighbourhood are, on average, older than 3 Gyr. Moreover, the signatures from planet engulfment are no longer discernible 3 Gyr after the occurrence if the engulfed mass is less than $10\ M_\oplus$ (assuming observable variations of the chemical composition exceed 0.02 dex). In this regard, we find these values to be smaller than proposed by SP21.


Conversely, a recent work by \cite{2024arXiv240313209L} also breaches this topic.
The authors report an occurrence rate of 8\% for engulfment events from a homogeneous sample of 91 co-natal pairs of stars with derived high-precision chemical abundances.
While SP21 attributed a difference in composition to be due to planet engulfment, \cite{2024arXiv240313209L} recognises that atomic diffusion can also impact the stellar composition. In this context, the authors tested two hypotheses: whether the observed variation in chemical composition is due to atomic diffusion or planet engulfment. 
Taking similar precautions as we did in this work, \cite{2024arXiv240313209L} obtained a result aligned with our findings, in particular for stars older than 3 Gyr.




The stellar age is also expected to have an impact in dictating how relevant the impact of planet engulfment would be in the chemical composition of the stellar surface, as the extension of the convective zone shrinks as the star evolves. Older stars would have thinner convective layers when compared with younger stars. However, since we did not have complete information on the stellar ages from an observational sample, we could not compared them with the outcome from simulations, hence  this was something we were unable to test.

Furthermore, the spectral type should also be taken into account, since for hotter stars not only is the dilution timescale shorter but the convective envelopes are shallower \cite[see e.g.][]{2018A&A...618A.132K, 2020A&A...643A.164S}, since higher temperatures require other transport mechanisms besides the traditional gravitational settling, which can influence the timescales of the mixing of the chemical imprint \citep{2006ARep...50.1001B, 2015ads..book.....M}. In this regard, it would be useful to test how much does the different temperatures of the binary components correspond to different sizes of the convective envelope, and thus how it would impact the dilution factor between both component of the binary system. Since in the sample of SP21 they consider only the mean effective temperature in their analysis and mention they allow differences up to 600 K, it would be interesting to verify what consequences this large difference could have in the results. In addition, a comparison with the sample from \cite{2024arXiv240313209L}, where they allow differences only up to 300 K, could provide further insight on the relevance of this effect.

Despite accounting for different aspects, we acknowledge our results have some limitations due to the nature  of the model they were based on. The 1000 stars simulated were Sun-like stars, thus the stellar models computed with Cesam2k20 consider chemical composition of the stellar atmospheres as solar-like. Moreover, when estimating the impact of planet engulfment on the stellar composition, we did not model each individual engulfment event for each system, even though the cumulative effect of multiple engulfment events could be greater.
The caveats for the Generation III Bern model are explained in detail in \cite{ngpps1}, and \cite{ngpps2} provides a comprehensive discussion on the limitations for the case of population synthesis itself. Consequences and constrains are further explored for different settings through the series of articles these works belong to \cite[e.g.][]{ngpps4,ngpps5,ngpps6}.

\begin{table}[]
\centering
\caption{Fraction of systems that register engulfment of planetary bodies with mass above a minimum value (2, 5 or 10 $M_\oplus$), during time intervals from a fixed point (60 Myr; 1, 2, 3, 4 or 5 Gyr) until 10 Gyr.}
\label{frac_systems}
\resizebox{0.8\columnwidth}{!}{%
\begin{tabular}{|c|c|c|}
\hline
\textbf{Time} & \textbf{$M_{planet,\ min}\ (M_\oplus)$}& \textbf{Fraction of systems (\%)} \\ \hline
\multirow{3}{*}{60 Myr}& 2  & 20.4\\
                        & 5  & 15.7\\
                        & 10 & 10.5\\ \hline
\multirow{3}{*}{1 Gyr}  & 2  & 16.8\\
                        & 5  & 12.9\\
                        & 10 & 7.3\\ \hline
\multirow{3}{*}{2 Gyr}  & 2  & 14.4\\
                        & 5  & 11.2\\
                        & 10 & 6.1\\ \hline
\multirow{3}{*}{3 Gyr}  & 2  & 11.9\\
                        & 5  & 8.9\\
                        & 10 & 4.7\\ \hline
\multirow{3}{*}{4 Gyr}  & 2  & 10.6\\
                        & 5  & 8.0\\
                        & 10 & 4.3  \\ \hline
\multirow{3}{*}{5 Gyr}  & 2  & 9.1\\
                        & 5  & 7.0\\
                        & 10 & 3.7  \\ \hline
\end{tabular}%
}
\end{table}

\section{Summary} \label{conclusions}

In this work we explore the impact of planet engulfment in the chemical composition of the host star. Using the results of Planetary Population synthesis from the Generation III Bern model \citep{ngpps1, ngpps2}, we were able to follow the formation and evolution of 1000 systems. In particular, the data allowed us to track the star and planets individually over time and analyse under which conditions planet engulfment might happen. 
An analysis of the timing of planet engulfment and amount of material involved reveals the existence of three different phases at different points of evolution, each with a main mechanism responsible for these occurrences. 

When comparing the properties of the systems with and without engulfment,  we found that systems that register an engulfment event at any point in time tend to form in more massive and more metallic disks. Under these conditions, giant planets are more likely to form, and though the presence of these planets can affect planet engulfment, they are not the sole reason behind these events. Through a simple model, we also determined the chemical composition of the planetary material, both the planets that were engulfed and the ones that remain in the system. We noted that, in general, the engulfed planets have a higher amount of refractories than the remaining planets, while the latter are richer in volatiles, in particular oxygen. This could hint to different formation places of these planets (inside and outside the ice-line).

In order to account for the physical processes happening in the stellar interior when determining the impact of planet engulfment on the stellar surface composition, we employ stellar models computed with Cesam2k20 \citep{morel08,marques13,deal18}. We simulate 9 cases of planets with varying masses being engulfed by their host star at different points in time and analyse the variation of the chemical composition in the stellar surface. Our results show that engulfment events happening while the protoplanetary disk is present leave no chemical imprint, as the convective layer ensures the total dilution of any material.
At later stages, when the disk is already dissipated, the variation on the chemical composition increases with time due to the convective layer shrinking as the star evolves. Moreover, since the engulfed planet will have higher masses as they had more time to evolve, the impact on the stellar chemical composition will likewise be greater. For instance, for events happening around 60 Myr, in order to still see a variation of $\sigma_{[Fe/H]}>0.02$ dex after 3 Gyr, we need an engulfment of a $M>10\ M_\oplus$ planet.

When looking at engulfment events that might generate an observable chemical imprint in the stellar surface and considering the majority of stars in the solar neighbourhood are older than 3 Gyr, we find an engulfment rate that cannot exceed $20\%$ for the optimistic case of being able to detect any variation above 0.02 dex in the stellar composition. Furthermore, if the detection threshold for planet engulfment is raised to 0.06 dex instead of 0.02 dex, this rate  will decrease even more. 

The stellar characterisation (e.g. chemical composition, mass, radius) is needed to infer information of orbiting planets, like their density or internal structure. Nonetheless, through engulfment events planets can also impact the stellar chemical composition, making the star-planet connection a bi-directional one.


\begin{acknowledgements}
      This work was financed by Portuguese funds through FCT – Fundação para a Ciência e a Tecnologia in the framework of the project 2022.06962.PTDC (EXO-Terra, DOI 10.54499/2022.06962.PTDC). 
      Co-funded by the European Union (ERC, FIERCE, 101052347). Views and opinions expressed are however those of the author(s) only and do not necessarily reflect those of the European Union or the European Research Council. Neither the European Union nor the granting authority can be held responsible for them. This work was supported by FCT - Fundação para a Ciência e a Tecnologia through national funds and by FEDER through COMPETE2020 - Programa Operacional Competitividade e Internacionalização by these grants: UIDB/04434/2020; UIDP/04434/2020.
      B.M.T.B.S is supported by an FCT fellowship, grant number 2022.11805.BD.
      M.D acknowledges support by CNES, focused on PLATO.
      S.G.S acknowledges the support from FCT through Investigador FCT contract nr. CEECIND/00826/2018 and  POPH/FSE (EC). 
      C.M. acknowledges support from the Swiss National Science Foundation under grant 200021\_204847 ``PlanetsInTime''. Parts of this work have been carried out within the framework of the NCCR PlanetS supported by the Swiss National Science Foundation under grants 51NF40\_182901 and 51NF40\_205606. 
      
\end{acknowledgements}

\bibliography{refs}
\bibliographystyle{aa}

\end{document}